\documentclass[aps, prb, twocolumn, nofootinbib, superscriptaddress, longbibliography,  floatfix]{revtex4-2}

\pdfoutput=1
\usepackage[utf8]{inputenc}
\DeclareUnicodeCharacter{00A0}{ }
\usepackage{amsmath}
\usepackage{amssymb}
\usepackage{physics}

\usepackage{graphicx}
\usepackage{color}
\usepackage[english]{babel}
\usepackage[nolist]{acronym}

\usepackage{hyperref}
\hypersetup{
	pdftoolbar=true,
	pdfmenubar=true,
	pdffitwindow=false,
	pdfstartview={FitH},
	pdftitle={},
	pdfauthor={},
	pdfsubject={},
	pdfcreator={},
	pdfproducer={},
	pdfnewwindow=true,
	colorlinks=true,
	linkcolor=black,
	citecolor=blue,
	filecolor=magenta,
	urlcolor=blue
}

\newcommand\nodag{{\vphantom{\dagger}}}
\newcommand\up{\uparrow}
\newcommand\dn{\downarrow}

\usepackage{xcolor}

\newcommand{\CCQ}{Center for Computational Quantum Physics, Flatiron Institute, 162 5th Avenue, New York, NY 10010, USA}
\newcommand{\CDF}{Coll\`ege de France, Universit\'e PSL, 11 place Marcelin Berthelot, 75005 Paris, France}
\newcommand{\CPHT}{CPHT, CNRS, Ecole Polytechnique, Institut Polytechnique de Paris, 91128 Palaiseau, France}
\newcommand{\saclay}{Universit\'e Paris-Saclay, CNRS, CEA, Institut de Physique Th\'eorique, 91191, Gif-sur-Yvette, France}

\begin{document}
\title{Algorithm for computing perturbation series of dynamical mean field theory}

\author{Corentin Bertrand}
\email{corentin.bertrand@eviden.com}
\affiliation{\CCQ}

\author{Michel Ferrero}
\affiliation{\CPHT}
\affiliation{\CDF}

\author{Olivier Parcollet}
\affiliation{\CCQ}
\affiliation{\saclay}

\date{\today}

\begin{abstract}
	\acresetall
We show how to use diagrammatic techniques 
to compute the weak-coupling perturbation series of the self-consistent solution to a \ac{DMFT} problem. 
This approach constitutes an alternative to 
using diagrammatic techniques directly as an impurity solver.
It allows one to bypass the need of multiple perturbative series resummations
within the \ac{DMFT} self-consistency loop. 
It can be applied at or out of equilibrium, with any diagrammatic formalism, such as real times, imaginary times, or Matsubara frequencies formalisms.
As a proof of principle, we
illustrate our method with the half-filled Hubbard model on the Bethe
lattice in the \ac{DMFT} approximation, using \ac{QQMC} to obtain the impurity perturbation series on the real time axis.
\end{abstract}

\maketitle

\begin{acronym}
	\acro{DMFT}{Dynamical Mean Field Theory}
	\acro{QQMC}{Quantum Quasi-Monte Carlo}
	\acro{DFT}{Discrete Fourier Transform}
	\acro{FTPS}{Fork Tensor Product State}
	\acro{NRG}{Numerical Renormalization Group}
	\acro{DCA}{Dynamical Cluster Approximation}
\end{acronym}

\section{Introduction}

\acresetall

Solving models of strongly correlated electrons on a lattice is notoriously
difficult, both analytically and numerically.  Embedding techniques, such as
\ac{DMFT}~\cite{Georges1996,KotliarRMP2006, Georges2016}
reduce the full lattice problem to a self-consistent quantum impurity problem, which
is significantly simpler to solve. 
At equilibrium, efficient numerical impurity solvers have been developed using
the imaginary time or Matsubara frequencies formalism, such as various 
continuous time quantum Monte Carlo algorithms~\cite{Gull2011}. 
However, there is an
increasing demand for solvers using a real time formalism, for two main
reasons.  First, experimentally accessible quantities, such as the spectral
function or the optical conductivity lie on the real frequency axis, and their
evaluation from Matsubara frequencies is an ill-conditioned
problem~\cite{Jarrell1996}, which requires extremely precise data~\cite{Fei2021}.
Second, non-equilibrium physics is only accessible from the real time axis, and
recent experimental developments have opened important questions in this field,
e.g. non-equilibrium phases transitions in strongly correlated
materials~\cite{Giannetti2016, Hedayat2021} or high temperature transport in
strange metals~\cite{Brown2019, Huang2019}.
While approximate real-time solvers have been used for embedding techniques for
some years~\cite{AokiWerner2014}, only a few of them are controlled, including
tensor network
techniques~\cite{Bauernfeind2017, CaoParcollet2023}, the inchworm
algorithm~\cite{Cohen2015, ErpenbeckCohen2022, GoldbergerCohen2024,
ErpenbeckGull2024} and real-time diagrammatic 
methods~\cite{ProfumoWaintal2015, Bertrand2019,Macek2020, Bertrand2021, NunezFernandezWaintal2022,
VanhoeckeSchiro2024}.

In this work, we focus on diagrammatic methods on the real time axis, and more specifically the weak coupling diagrammatic expansion~\cite{Rubtsov2004}.
Perturbative techniques have recently made remarkable progresses  in this area, 
in two directions: the precise computation of relatively high perturbation orders~\cite{ProfumoWaintal2015, Macek2020,Bertrand2021,
NunezFernandezWaintal2022}, 
and methods to sum the perturbative series into strong coupling regimes, beyond their weak coupling 
radius of convergence~\cite{ProfumoWaintal2015, Bertrand2019}.
Tensor networks (tensor cross interpolation) are for example now able to compute up to 30 perturbation orders
for a simple quantum dot model, 
with a much faster convergence than previous Monte Carlo, or quasi-Monte Carlo techniques~\cite{NunezFernandezWaintal2022}.
Many approaches have been discussed to sum the series, see e.g. conformal
transformations~\cite{ProfumoWaintal2015, Bertrand2019, Simkovic2019} or Padé
approximants~\cite{Baker1996, Bertrand2021} and
generalizations~\cite{Simkovic2019}, or the more recent cross-extrapolation
technique~\cite{JeanninWaintal2024}.
While they are in some cases very successful~\cite{Bertrand2019, JeanninWaintal2024}, 
they have not yet been made robust enough to make a black box algorithm
suitable for use in a self-consistent DMFT loop.
As a result, the application of
real-time axis diagrammatic techniques as DMFT impurity solvers has proven difficult until now.

In this work, we propose to address this issue from another angle. We
present a simple algorithm to compute the perturbation series of the
DMFT \emph{self-consistent solution} order by order, instead of 
using the diagrammatic approach just as an impurity solver for a fixed bath.
Given any weak coupling diagrammatic technique for the quantum impurity model, 
we show how to extend it to simultaneously solve the impurity model
and the DMFT self-consistency condition, order by order.
This brings several advantages.  
First, we only need to perform one resummation of the perturbative series, 
eliminating the need for a series resummation at each DMFT iteration.
Second, it allows for a direct study of the analytical structure of the
self-consistent solution, which contains for instance information on phase
transitions through the location of singularities.
We emphasize that our approach is independent on the formalism used for the diagrammatic expansion (Matsubara, imaginary times, Keldysh, etc), so that it is applicable to equilibrium, non-equilibrium steady states as well as transient states such as quenches.

This paper is organized as follows.
Section~\ref{sec:dmft_equations} recalls the basic DMFT equations and some notations.
Sec.~\ref{sec:algorithm} describes the algorithm.
As a proof of principle, we then illustrate it on the Hubbard model in the Bethe lattice in equilibrium in Sec.~\ref{sec:illustration}.
Finally we conclude in Sec.~\ref{sec:conclusion}.

\section{DMFT equations}
\label{sec:dmft_equations}

We consider a DMFT problem,  described as the combination of an
impurity problem and a self-consistency relation between the impurity and its
bath. We denote by $U$ the interaction on the impurity, 
by $G$ the Green function of the impurity, and $\mathcal{G}$ the Weiss field~\cite{Georges1996}, i.e. the Green
function of the impurity for $U=0$.
The impurity self-energy $\Sigma$ is defined from the
Dyson equation $G = \mathcal{G} + \mathcal{G} \Sigma G$.
We will also use the improved estimator $F \equiv \Sigma G$~\cite{Bulla2001}.
In full generality, these quantities ($G$, $\mathcal{G}$, $F$ and $\Sigma$)
are matrix-valued functions of two times on the Keldysh contour, and products are understood as convolutions on the Keldysh contour.
This can be specialized to, e.g. real frequencies in case of a steady state problem, or Matsubara frequencies in equilibrium problems.

We assume that we have an impurity solver that, given the Weiss field
$\mathcal{G}$, produces the perturbative expansion of $G$, $F$ or $\Sigma$. To
be specific, we will assume here that it produces the improved estimator $F$.
We can use e.g. Monte Carlo~\cite{ProfumoWaintal2015, Bertrand2019, Bertrand2021}, quasi-Monte Carlo~\cite{Macek2020} or tensor cross interpolation
~\cite{NunezFernandezWaintal2022}. In this work, we will use quasi-Monte Carlo.

The improved estimator $F$ is a functional of $\mathcal{G}$ and a series in $U$.
It is the sum of Feynman diagrams with each vertex contributing a factor $U$ and with the bare propagator $\mathcal{G}$.
Grouping the diagrams with the same number $n$ of vertices into functionals $\mathcal{I}_n[\mathcal{G}]$ yields
\begin{equation}
	\label{eq:impurity_solver}
	F = \sum_{n \ge 1} U^n \mathcal{I}_n[\mathcal{G}]
\end{equation}
Note that $\mathcal{I}_0 = 0$, since $F$ has no order zero contribution by definition (the same is true for the self-energy).
This will be crucial in the following.

The DMFT self-consistency condition relates the Weiss field to the impurity $F$ (or the self-energy $\Sigma$).
It can be written as a functional $\mathcal{S}$ such that
\begin{equation}
	\label{eq:sc_condition}
	\mathcal{G} = \mathcal{S}\qty[F].
\end{equation}
$\mathcal{S}$ depends on the details of the non-interacting model.

The system of equations formed by~\eqref{eq:impurity_solver} and~\eqref{eq:sc_condition} defines the DMFT problem to solve.
We wish to solve this problem order-by-order in $U$, i.e. to compute the perturbation series for $F(U)$ and $\mathcal{G}(U)$:
\begin{subequations}
	\begin{gather}
		\mathcal{G}(U) = \sum_{n \ge 0} U^n \mathcal{G}_n
		\\
		F(U) = \sum_{n \ge 1} U^n F_n
	\end{gather}
\end{subequations}
From $F$ and $\mathcal{G}$, the impurity Green function is obtained using $G(U) = \mathcal{G}(U)(F(U)+1)$.
We also define the partial sums:
\begin{subequations}
\begin{gather}
	\mathcal{G}^{(m)}(U) \equiv \sum_{n = 0}^m U^n \mathcal{G}_n
	\\
	F^{(m)}(U) \equiv \sum_{n = 1}^m U^n F_n
\end{gather}
\end{subequations}
Throughout the rest of the paper we will use $[\ldots]_m$ to denote the order $m$ coefficient of the series between brackets.

\section{Algorithm}
\label{sec:algorithm}

We \emph{simultaneously} compute the perturbation series
of $F$ and $\mathcal{G}$, order by order, starting from the
knowledge of $\mathcal{G}_0$, up to some order $N$.
Our method directly produces the exact perturbative solution of the self-consistent DMFT solution:
there is no iteration in the usual DMFT sense.

\subsection{Mathematical preliminaries}
\label{sec:coeff_extraction}

Our method relies on the ability to extract the coefficient of a polynomial of
known maximum degree from its evaluations on roots of unity in the complex plane.

Consider a polynomial $P$ of degree at most $d$
\begin{equation}
	P(U) = \sum_{k=0}^d P_k U^k
\end{equation}
and define the root of unity
\begin{equation}
	\xi_d = \exp(\frac{2i\pi}{d+1}).
\end{equation}
The coefficients $P_k$ of $P$ are related to the evaluation of $P$ on the $d+1$ roots of unity $1, \xi_d, \xi_d^2, \ldots, \xi_d^d$ by an inverse \ac{DFT}:
\begin{equation}
	P_k = \frac{1}{d+1}\sum_{l=0}^{d} P(\xi_{d}^l) \; \xi_{d}^{-kl}
\end{equation}
For numerical stability, it may be advantageous to evaluate $P$ at points with a modulus lower or higher than one.
This is done by rescaling the complex plane. Introducing a scaling factor $r > 0$, observe that $P(U)$ is also a polynomial in the variable $V = U/r$ with coefficients $r^k P_k$.
Using the same relation on the transformed polynomial leads to
\begin{equation}
	\label{eq:roots_unity_trick}
	P_k = \frac{r^{-k}}{(d+1)}\sum_{l=0}^{d} P(r \xi_{d}^l) \; \xi_{d}^{-kl}.
\end{equation}
Here, $P$ is evaluated on the circle of radius $r$.

\subsection{Algorithm}

We work recursively on the order $N$ of the expansion of $\mathcal{G}$ and $F$.  At step $N$,
we know $\mathcal{G}_0, \ldots, \mathcal{G}_{N-1}$ and $F_1, \ldots,
F_{N-1}$ and we compute $F_N$ and $\mathcal{G}_N$.

\subsubsection{Computing $F_N$}
An expression for $F_N$ is obtained by extracting the order $N$ coefficient of the right-hand-side of Eq.~\eqref{eq:impurity_solver}, term-by-term, and keeping in mind that $\mathcal{G}$ is the self-consistent Weiss field which depends on $U$.
Because of the factor $U^n$, terms with $n > N$ do not contribute.
The other terms contribute to the order $N-n$ coefficient of $\mathcal{I}_n[\mathcal{G}(U)]$, which means
\begin{equation}
	\label{eq:decomp_K_N}
	F_N = \sum_{n=1}^N \qty[\mathcal{I}_n[\mathcal{G}(U)]]_{N-n}.
\end{equation}
Since $\mathcal{I}_n[\mathcal{G}(U)]$ is a sum of diagrams of finite order, it is a polynomial in the propagator $\mathcal{G}(U)$ whose coefficients are independent of $U$.
As such, once seen as a polynomial in $U$, its coefficient in $U^{N-n}$ cannot involve $\mathcal{G}_k$ for $k > N-n$.
As a result, we can replace $\mathcal{G}(U)$ by its partial sum $\mathcal{G}^{(N-n)}(U)$:
\begin{gather}
	\label{eq:impurity_solver_with_approx_g}
	\qty[\mathcal{I}_n[\mathcal{G}(U)]]_{N-n} = \qty[\mathcal{I}_n\qty[\mathcal{G}^{(N-n)}(U)]]_{N-n}.
\end{gather}
This shows that $F_N$ can be obtained from the knowledge of $\mathcal{G}_0, \ldots, \mathcal{G}_{N-1}$ only.
Crucially, $F_N$ does not depend on $\mathcal{G}_N$, as its diagrams contain at least one vertex (the same is true for the self-energy).

The right-hand side of Eq.~\eqref{eq:impurity_solver_with_approx_g} is
evaluated using the roots of unity formula Eq.~\eqref{eq:roots_unity_trick}.
$\mathcal{I}_n\qty[\mathcal{G}^{(N-n)}(U)]$ is indeed a polynomial in $U$
(since it is a polynomial in the propagator $\mathcal{G}^{(N-n)}(U)$, which
itself is a polynomial in $U$).  Moreover, order $n$  diagrams for $F$ have $2n$
propagators\footnote{Assuming the interaction vertex is four-legged.}, and each
contributes a power in $U$ up to $N-n$. So
$\mathcal{I}_n\qty[\mathcal{G}^{(N-n)}(U)]$ is a polynomial in $U$ of degree at most
$d_{n,N} \equiv 2n(N-n)$.

Each term of Eq.~\eqref{eq:decomp_K_N} can therefore be obtained by applying Eq.~\eqref{eq:roots_unity_trick} with the corresponding degree.
This gives an explicit formula for $F_N$:
\begin{equation}
	\label{eq:coeff_extraction_1}
	F_N = \sum_{n=1}^N\frac{r^{-(N-n)}}{d_{n,N}+1}\sum_{l=0}^{d_{n,N}} \mathcal{I}_n\qty[\mathcal{G}^{(N-n)}(r\xi_{d_{n,N}}^l)] \; \xi_{d_{n,N}}^{-(N-n)l}
\end{equation}
for any choice of $r > 0$.
Let us stress that the Weiss field is a polynomial in $U$, so 
its evaluation at complex $U$ is well defined.
The root of unity method avoids the need of enumerating all diagrams contributing to
$\mathcal{I}_n$, and then looking up all choices of propagators among
$\{\mathcal{G}_0, \ldots, U^{N-n} \mathcal{G}_{N-n}\}$ that produce a
contribution proportional to $U^{N-n}$.  Here, we perform this selection
analytically, without enumerating diagrams, in a way that is independent on how
$\mathcal{I}_n$ is computed (enumerating diagrams explicitly, or using determinants algorithms).

Finally, let us emphasize that Eq.~\eqref{eq:coeff_extraction_1} has the same form in a real (respectively Matsubara) frequency formalism, if the impurity solver works in the steady state (resp. equilibrium) with real (resp. Matsubara) frequencies.
This is because Eq.~\eqref{eq:coeff_extraction_1} executes a coefficient extraction, which is a linear operation, and thus commutes with basis transform (e.g. the Fourier transform).

\subsubsection{Computing $\mathcal{G}_N$}
\label{sec:self-consistency}

We now compute $\mathcal{G}_N$ from the self-consistency condition Eq.~\eqref{eq:sc_condition}.
Since the form of the self-consistency condition $\mathcal{S}$ depends on the DMFT problem under study, each case should be derived separately.

However, we will assume that the self-consistency condition does not depend on $U$, and $\mathcal{S}[F]$ is analytic around $F=0$, such that we have
\begin{equation}
	\label{eq:sc_condition_obo}
	\mathcal{G}_N = \qty[\mathcal{S}\qty[F^{(N)}]]_N.
\end{equation}
We are not aware of \ac{DMFT} problems that break these assumptions, at or out of equilibrium.
Therefore, computing $\mathcal{G}_N$ requires only the knowledge of $F_1, \ldots, F_N$, which we obtained previously.

We now give a more precise formulation in two usual cases: a Bravais
lattice and the Bethe lattice, limiting ourselves to single-site DMFT for simplicity.

On a Bravais lattice, the non-interacting case is described by a momentum (denoted $\vb{k}$) dependent Green function $g_0(\vb{k})$, which is easily computed.
The self-consistency condition can then be written as the set of equations
\begin{subequations}
	\begin{gather}
		\label{eq:break_down_1}
		G = \mathcal{G} + G \Sigma \mathcal{G},
		\\
		\label{eq:break_down_2}
		G = \int_{\rm BZ} \dd{\vb{k}} g(\vb{k}),
		\\
		\label{eq:break_down_3}
		g(\vb{k}) = g_0(\vb{k}) + g(\vb{k}) \Sigma g_0(\vb{k}),
	\end{gather}
\end{subequations}
where $g(\vb{k})$ is the interacting lattice Green function at momentum $\vb{k}$, and $\int_{\rm BZ}$ denotes the normalized integral over the first Brillouin Zone.
Let us assume that we already have $\mathcal{G}$, $G$, $\Sigma$ and $g$ up to order $N-1$, as well as $F$ up to order $N$.
$\Sigma_N$ is then computed by developing $\Sigma = F G^{-1}$, using $G_0 = \mathcal{G}_0$:
\begin{equation}
	\Sigma_N = \qty(F_N - \sum_{k=1}^{N-1} \Sigma_k G_{N-k}) \mathcal{G}_0^{-1}
\end{equation}
Using Eq.~\eqref{eq:break_down_3}, we get
\begin{equation}
	g_N(\vb{k}) = \sum_{k=0}^{N-1} g_k(\vb{k}) \Sigma_{N-k} g_0(\vb{k})
\end{equation}
From Eq.~\eqref{eq:break_down_2} we then obtain
\begin{equation}
	G_N = \int_{\rm BZ} \dd{\vb{k}} \sum_{k=0}^{N-1} g_k(\vb{k}) \Sigma_{N-k} g_0(\vb{k})
\end{equation}
We now have $\Sigma$ and $G$ up to order $N$, which finally gives $\mathcal{G}_N$ using Eq.~\eqref{eq:break_down_1}:
\begin{equation}
	\mathcal{G}_N = G_N - \sum_{k=0}^{N-1} \sum_{l=1}^{N-k} G_{N-k-l} \Sigma_l \mathcal{G}_k.
\end{equation}

On the Bethe lattice, we show in App.~\ref{app:sc_bethe_lattice} that
\begin{equation}
	\mathcal{G}_N - t^2 G_0 \mathcal{G}_N G_0 = t^2 G_0 \qty[\mathcal{G}^{(N-1)} \qty(F^{(N)} + 1) \mathcal{G}^{(N-1)}]_N
\end{equation}
It allows for the computation of $\mathcal{G}_N$ from the knowledge of $\mathcal{G}$ up to order $N-1$ and $F$ up to order $N$, by inverting the linear operator $x \rightarrow x - t^2 G_0 x G_0$.

In a steady state problem, one can go to the frequency domain, leading to the same formula where $\mathcal{G}$, $G$ and $F$ are replaced by $2\times2$ matrices containing the retarded, advanced and Keldysh components~\cite{Kamenev_2011}, e.g.
\begin{equation}
	\mathcal{G} \rightarrow
	\begin{pmatrix}
		\mathcal{G}^R(\omega) & 0
		\\
		\mathcal{G}^K(\omega) & \mathcal{G}^A(\omega)
	\end{pmatrix}
\end{equation}
where $\mathcal{G}^R(\omega)$, $\mathcal{G}^A(\omega)$ and $\mathcal{G}^K(\omega)$ are respectively the retarded, advanced and Keldysh components of the Weiss field at frequency $\omega$.
In the following, retarded components are denoted with a $R$ superscript.

In particular, the self-consistency for the retarded component at frequency $\omega$, relevant for equilibrium problems, simplifies into
\begin{equation}
	\label{eq:sc_bethe_obo}
	\mathcal{G}_N^R = \frac{t^2}{\sqrt{(\omega + \mu)^2 - 4t^2}}\qty[\mathcal{G}^{R(N-1)} \qty(1 + F^{R(N)}) \mathcal{G}^{R(N-1)}]_N,
\end{equation}
where $\mu$ is a chemical potential.
This is shown in App.~\ref{app:sc_bethe_lattice}.

\subsection{Computational cost}

Most of the computational cost of the algorithm originates from $\mathcal{I}_n$, i.e. the perturbative solution of the impurity problem.
For a standalone impurity problem, obtaining the  perturbation series for $F$ requires one call to $\mathcal{I}_n$ for each order $n$.
Let us now compute the number of calls to $\mathcal{I}_n$ needed to obtain the series of DMFT solution for $F$.

\begin{figure}[t]
	\centering
	\includegraphics[width=\linewidth]{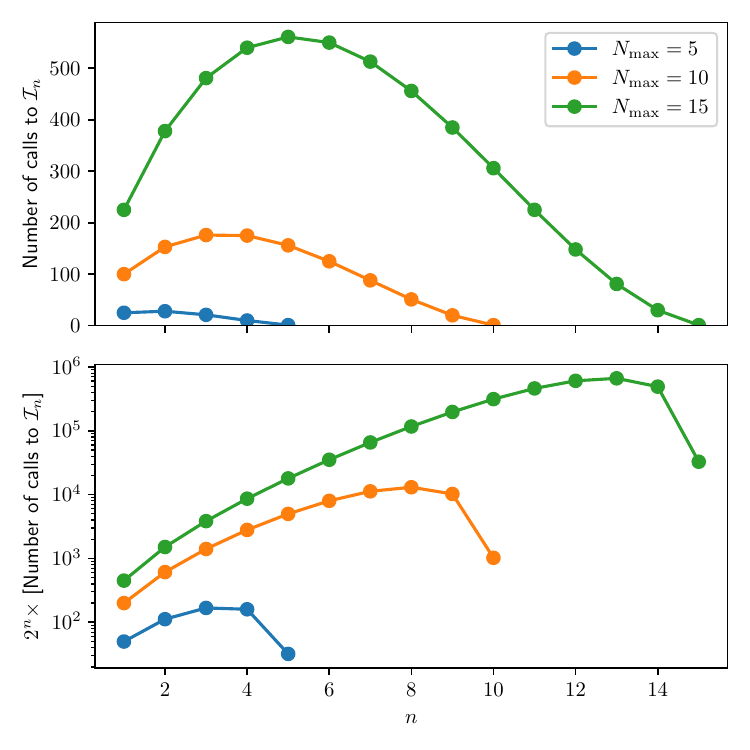}
	\caption{
		\label{fig:cost}
		Upper panel: number of calls to the impurity solver at perturbation order $n$ so as to compute the DMFT solution up to order $N_{\rm max}$ (Eq.~\eqref{eq:cost}).
		Lower panel: computational cost assuming a call to $\mathcal{I}_n$ costs $2^n$.
		Most of the cost lies in the last few orders. The overall cost is exponential in $N_{\rm max}$, as expected.
	}
\end{figure}

During the $N$th step of our algorithm (i.e. the step computing $\mathcal{G}_N$ and $F_N$), calls to every $\mathcal{I}_n$ for $n = 1,\ldots, N$ must be made.
For a given $n$, $\mathcal{I}_n$ is called $d_{n,N} + 1 = 2n(N-n) + 1$ times.
Cumulating over all steps up to $N_{\rm max}$, $\mathcal{I}_n$ must be called a total number of times
\begin{equation}
	\label{eq:cost}
	\begin{split}
		\sum_{N=n}^{N_{\rm max}} \qty[2n(N-n) + 1] = \qty(1 + n(N_{\rm max} - n))(N_{\rm max} - n + 1)
	\end{split}
\end{equation}
This number of calls is illustrated in Fig.~\ref{fig:cost} upper panel.
The total number of calls to $\mathcal{I}_n$, for a fixed $n$, is thus quadratic in $N_{\rm max}$.

It is expected that a call to $\mathcal{I}_n$ has a complexity in time that increases exponentially with $n$, as is the case for example using \ac{QQMC}~\cite{Macek2020} or CDet~\cite{Rossi2017}.
We illustrate the computational cost, making the simplified assumption that a call to $\mathcal{I}_n$ costs $2^n$, and ignoring any potential parallelization, in Fig.~\ref{fig:cost} lower panel.
The overall cost is exponential in $N_{\rm max}$, as expected.

Surprisingly, the computation time is not dominated by the last order, but by the last few orders before it.
Indeed $\mathcal{I}_{N_{\rm max}}$ is called only once, 
because diagrams of order $N_{\rm max}$ with bare propagators $\mathcal{G}_0$ already contribute to the power $U^{N_{\rm max}}$, so that higher orders in the propagator are not needed.
The coefficient extraction is then trivial and a single root of unity is used.

Many calls to $\mathcal{I}_n$ are independent. Indeed, evaluation over each set of roots of unity form a batch of independent calculations.
Therefore, there is an opportunity for parallelization, especially in the case the impurity solver is not easily made parallel.
In our illustration of Sec.~\ref{sec:illustration} we use \ac{QQMC}, which is already parallel, so we do not make use of this opportunity.

\subsection{Mitigating the sign/phase problem}

In the real or imaginary times formalism, computing $\mathcal{I}_n$ with diagrammatic Monte Carlo generally implies high dimensional integrals.
For example, algorithms recently developed for the real-time Keldysh formalism~\cite{NunezFernandezWaintal2022, ProfumoWaintal2015, Bertrand2019, Macek2020} (including \ac{QQMC}~\cite{Macek2020} which we employ in Sec.~\ref{sec:illustration}) compute $\mathcal{I}_n$ by summing an $n$-dimensional integral:
\begin{equation}
	\label{eq:QQMC_integral}
	\mathcal{I}_n[\mathcal{G}] = \int \dd{\vb{u}} \mathcal{F}_n[\mathcal{G}, \vb{u}]
\end{equation}
Here the integral runs over the hypercube $[0, t_{\rm M}]^n$ with a large $t_{\rm M}$.
Definitions of $\mathcal{F}_n$ can be found in Refs.~\cite{Bertrand2019a, Bertrand2021} for example.

It is a legitimate concern that the introduction of complex roots of unity and nonphysical Weiss functions may bring an additional sign or phase problem to the evaluation of such integrals, when compared to the standard impurity problem.
Also, large cancellations may occur when taking the inverse \ac{DFT}, bringing in more difficulties to its numerical evaluation.

These problems can be mitigated by performing the \ac{DFT} under the integral.
This means that instead of extracting the coefficients of the integral $\mathcal{I}_n[\mathcal{G}(U)]$, we extract those of the integrand $\mathcal{F}_n[\mathcal{G}(U), \vb{u}]$.
This way, the integrand is the same as if we selected the correct diagrams from an exhaustive enumeration, so that no additional sign/phase problem is introduced by the coefficient extraction technique.
This does not reduce the inherent sign/phase problem of the impurity model.

\section{Illustration on the Bethe lattice}
\label{sec:illustration}

As a proof of principle, we illustrate our algorithm on the DMFT solution of the 
half-filled equilibrium Hubbard model on the Bethe lattice.
We work with the Keldysh formalism, obtaining Green's functions directly on the real frequency axis.

\subsection{Model}

We consider the single band Fermi-Hubbard model on the Bethe lattice, whose Hamiltonian is
\begin{subequations}
\begin{gather}
	\label{eq:Hamiltonian}
	H = H_0 + U H_{\rm int}
	\\
	H_0 = -\mu \sum_i \sum_{\sigma} n_{i,\sigma}
	- t \sum_{\langle i,j \rangle} \sum_{\sigma} c^\dag_{i,\sigma} c^\nodag_{j,\sigma}
	\\
	H_{\rm int} = \sum_i (n_{i\up} - 1/2)(n_{i\dn} - 1/2)
\end{gather}
\end{subequations}
where $c^\nodag_{i,\sigma}$ and $c^\dag_{i,\sigma}$ are respectively the annihilation and creation operators for an electron of spin $\sigma$ on site $i$, $n_{i,\sigma} = c^\dag_{i,\sigma} c^\nodag_{i,\sigma}$ is the number operator, $\mu = 0$ is the chemical potential at half-filling, $t$ the hopping energy, $\langle i,j \rangle$ denotes all pairs of neighboring sites.
$U$ is the perturbative parameter.
We work in unit of the hopping i.e. $t=1$.

In the non-interacting case, the self-consistency condition leads to a semi-circular density of states with half-bandwidth $D = 2t$.
In order to avoid the numerical issue of having singular band edges we add a weak dissipation term $\eta = t \times 10^{-2}$ on every site of the lattice.
Equivalently, we add a small imaginary self-energy to the chemical potential,  $\mu \rightarrow \mu + i\eta$, 
which gives the following non-interacting retarded Weiss function~\cite{Georges1996}:
\begin{equation}
	\mathcal{G}_0^R(\omega) = 
	\begin{cases}
		\omega + i\eta + \sqrt{(\omega + i\eta)^2 - D^2},   & \omega < -D,
		\\
		\omega + i\eta - i \sqrt{D^2 - (\omega + i\eta)^2}, & |\omega| \leq D,
		\\
		\omega + i\eta - \sqrt{(\omega + i\eta)^2 - D^2},   & \omega > D.
	\end{cases}
\end{equation}
The other (Keldysh) components are obtained from  equilibrium fluctuation-dissipation relations.

\subsection{Results: perturbative series}

We apply our algorithm to compute the perturbation series of the corresponding single-site \ac{DMFT} problem.
We work in real frequencies and we use \ac{QQMC}~\cite{Bertrand2021} to compute $\mathcal{I}_n$ and obtain the perturbation series of the retarded component of $F$.
We apply the self-consistency condition between $F$ and the Weiss function from Eq.~\eqref{eq:sc_bethe_obo}.
Details on the implementation and error estimation can be found in App.~\ref{app:implementation}.

\begin{figure}[t]
	\centering
	\includegraphics[width=\linewidth]{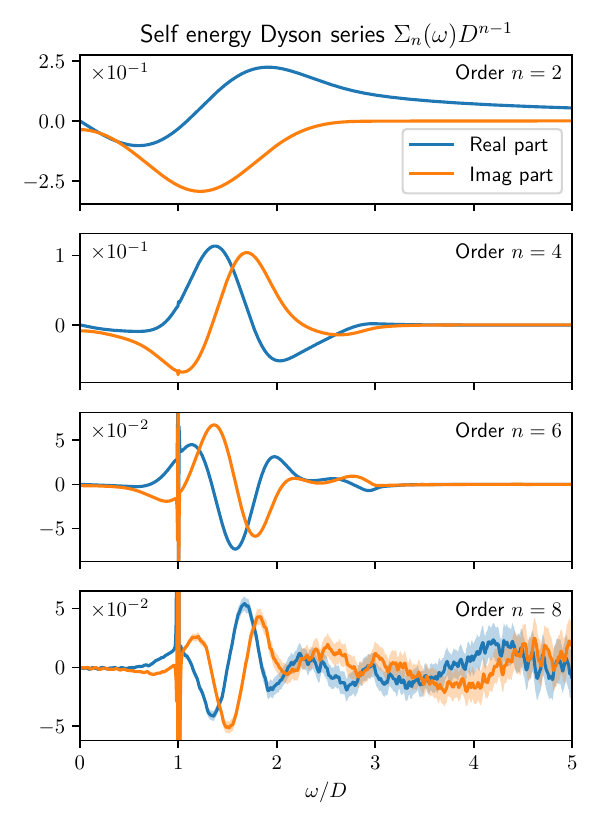}
	\caption{
		\label{fig:self_energy}
		Perturbation series of the \ac{DMFT} self-energy for the Hubbard model on the Bethe lattice at half filling. Negative frequencies can be deduced from particle-hole symmetry and odd orders are zero. Shaded areas are error bars. The singular behavior at $\omega = D$ is a numerical artifact (see main text).
		Temperature is $T=0.1D$.
	}
\end{figure}

We first check that the two first perturbative orders we obtain match known analytical values at zero temperature~\cite{GebhardNoack_2003}.
The data can be seen in App.~\ref{app:comp_exact}.

Then, we show in Fig.~\ref{fig:self_energy} the perturbation series of the retarded component of the self-energy of the DMFT solution, up to order 8.
The temperature is $T=0.1D$.
The shaded areas represent integration error bars, using the standard randomized quasi-Monte Carlo approach~\cite{Dick2013, LEcuyer2018, Macek2020}.

The sharp features appearing at the band edge $\omega = D$ are numerical artifacts that we explain as follows.
In the $\eta=0$ case, the non-interacting Green function $\mathcal{G}^R_0$ has singularities at the band edges. At small $U > 0$ however, the Green function $G^R$ and the Weiss function $\mathcal{G}^R$ are smooth in $\omega$. Their perturbation coefficients must then have singularities at the band edges.
On the other hand, the self-energy is zero at $U=0$ and smooth at small $U > 0$, so its perturbation series has no such singularities.
As a result, in the Dyson equation the singularities of $G^R$ and $\mathcal{G}^R$ must cancel each other. The band edge features of Fig.~\ref{fig:self_energy} are numerical inaccuracies in this cancellation between very large values.
For a small $\eta > 0$, singularities are replaced with high peaks, and similar cancellations between large values occurs.

\subsection{Comparison to equilibrium DMFT solver}

\begin{figure}[t]
	\centering
	\includegraphics[width=\linewidth]{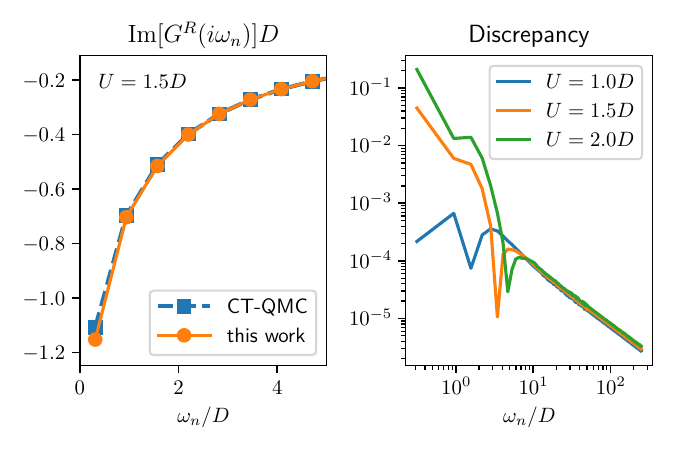}
	\caption{
		\label{fig:matsubara}
		Comparison to standard DMFT calculation on Matsubara frequencies, using CT-HYB impurity solver.
		(left) Matsubara Green function at $U=1.5D$ with standard DMFT loop (squares) and using the resummation of the order 6 perturbation series (circles).
		Due to particle-hole symmetry, the real part is zero.
		(right) Discrepancy between the two methods on a range of $U$.
		Temperature is $T=0.1D$.
	}
\end{figure}

We compare our real-time axis computation to a standard Matsubara frequencies DMFT loop computation performed with a quantum Monte-Carlo impurity solver, namely the CT-HYB algorithm~\cite{Werner2006} implemented with the TRIQS library~\cite{Parcollet2015}.
Starting from the computation of $G$ up to order 6 with our method, we obtain the perturbation series for the Matsubara Green function, defined as
\begin{eqnarray}
	G^R(i\omega_n) = \int \dd{\omega} \frac{A(\omega)}{i\omega_n - \omega}
\end{eqnarray}
where $\omega_n = (2n+1)\pi T$ are Matsubara frequencies, and $A(\omega)$ is the local spectral function on the lattice.
The series, considered as a function of $U^2$, is then resummed at given $U^2$ with Pad\'e approximants.
More precisely, we take the median of all Pad\'e approximants that we can compute, i.e. those of order $(k, l)$ with $k+l \le 3$.
We checked that variance from the \ac{QQMC} solver only weakly affects the result.

The comparison is shown in Fig.~\ref{fig:matsubara}.
On the left panel, we show the imaginary part of the Matsubara Green function (real part is zero by particle-hole symmetry), at $U=1.5D$.
We see a good agreement with CT-HYB, and we attribute the small discrepancy on the first Matsubara frequency to resummation error.
The absolute difference is shown on the right panel, for a range of values for $U$.
One can see that at low Matsubara frequencies, the discrepancy increases as we are getting closer to the phase transition ($U_{\rm c2} \approx 2.3D$), which is expected due to the presence of a singularity near or at the transition.
At high Matsubara frequencies, the error falls off as $1/\omega_n$ and weakly depends on $U$, indicating that resummation is not affected by the phase transition at these frequencies.

\subsection{Metallic solution}

\begin{figure}[t]
	\centering
	\includegraphics[width=\linewidth]{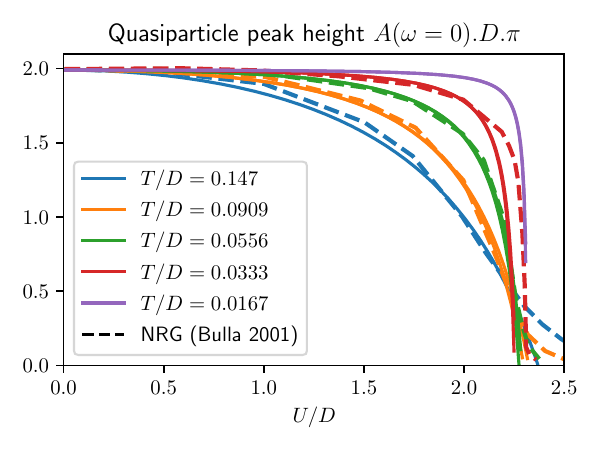}
	\caption{
		\label{fig:quasiparticle_height}
		Height of the quasiparticle peak as a function of interaction strength (plain lines)  compared to NRG results from Ref.~\cite{Bulla2001} (dashed lines).
		Even with as few as 6 orders of perturbation, the critical $U$ at which the quasiparticle peak vanishes can be located within an error bar $\sim 0.1 D$.
		The two lowest temperatures (red and purple) are below the critical temperature.
	}
\end{figure}

We resum the perturbative expansion, starting from $U=0$, 
in the metallic DMFT solution, and we compare our result to a numerically exact 
\ac{NRG} solution of Bulla et al.~\cite{Bulla2001}.

In Figure~\ref{fig:quasiparticle_height}, we present
the quasiparticle peak height, defined as the spectral function at the Fermi level:
\begin{equation}
	A(\omega=0) = -\frac{1}{\pi} \Im[G^R(\omega=0)]
\end{equation}
as a function of $U/D$.
The series for $A(\omega=0)$ was computed up to order $6$ and resummed using Padé approximants~\cite{Baker1996}.
We present different temperatures, above (blue, orange and green lines) and below (red and purple lines) the DMFT critical temperature $T_{\rm c} \approx 0.04 D$, and compare them to the \ac{NRG} calculation~\cite{Bulla2001}.

Above $T_{\rm c}$, as $U$ increases, the quasiparticle peak height drops to zero. The inflection point locates the crossover region between metal and insulator.
Our results match well the \ac{NRG} calculations until the crossover region. There, the resummed series keep the same curvature and end up changing sign.
The point where $A(\omega=0)$ becomes negative give an approximate location of the crossover region, accurate within $\sim 0.1 D$.
Below $T_{\rm c}$, the crossover becomes a phase transition, and the quasiparticle height should vanish at $U = U_{\rm c2}(T)$. Our result is still accurate below the transition, and gives the location of $U_{\rm c2}$ with an error of about $0.1D$ (red line).

This benchmark shows that, even with only 6 orders in perturbation theory (note that odd orders are zero by symmetry), 
the direct perturbative expansion of the DMFT solution, when properly resummed, 
has the potential to describe the metallic solution, up to $U\approx U_{c2}$, and the related finite temperature crossovers.
We expect that the precision could be significantly improved by 
using our  method with more sophisticated perturbative impurity solver \cite{NunezFernandezWaintal2022}, 
which would yield more perturbative orders and a higher precision.

\section{Conclusion}
\label{sec:conclusion}

We presented a simple algorithm to extend any weak coupling diagrammatic technique for quantum impurity model 
to directly compute, order by order,
the perturbative expansion of a DMFT solution, by expanding
simultaneously the solution of the impurity solver and the self-consistency
condition.
Our approach constitutes an alternative to the straightforward application 
of diagrammatic technique to the impurity solver, at fixed bath.
It avoids the need of resummation of the perturbative series within the DMFT 
self-consistency iterations.
The approach is independent of the type of weak coupling diagrammatic impurity solver used, 
be it in real
times, real or Matsubara frequencies, at or out of equilibrium.  It can also be
applied directly to several flavors of embedding techniques, such as cluster
DMFT or the \ac{DCA}~\cite{MaierHettler2005}.
We benchmarked it on an explicit computation of a standard DMFT solution on the Bethe lattice in equilibrium.

This work is mainly a proof of concept, and its applications still have to be developed further, 
for example for out-of-equilibrium solutions of DMFT for which solvers are much less developed than their equilibrium counterparts.
At high temperature, it could 
provides a controlled way to study crossover lines above quantum critical
points, or the problem of transport in high temperature strange
metals~\cite{Brown2019, Huang2019}.  Indeed, the physics far from a phase
transition is a natural target for diagrammatic techniques.
Furthermore, the analytical structure of the solution of an embedding theory can 
be studied directly from the perturbative expansion. In particular, the location of singularities close to or
on the real axis of $U$ provides indications about crossovers or phase transitions, even
with a low number of orders.


\begin{acknowledgments}
The Flatiron Institute is a division of the Simons Foundation.

This work was granted access to the HPC resources of TGCC and IDRIS under the allocations A0170510609 attributed by GENCI (Grand Equipement National de Calcul Intensif).
\end{acknowledgments}

\appendix

\section{Order-by-order self-consistency condition on the Bethe lattice}
\label{app:sc_bethe_lattice}

We denote $g$ the Green function of an isolated site, i.e. uncoupled to the rest of the lattice.
Then, the non-interacting Green function on any site of the Bethe lattice, denoted $G_0$, respects the Dyson equation
\begin{equation}
	\label{eq:bethe_lattice_dyson}
	G_0 = g + t^2 G_0^2 g,
\end{equation}
with $t$ the hopping amplitude.
In the presence of interactions, since the Green function on each site is $G$, the Weiss field $\mathcal{G}$ respects another Dyson equation,
\begin{equation}
	\mathcal{G} = g + t^2 g G \mathcal{G}.
\end{equation}

Multiplying this last equation on the left by $(1 + t^2 G_0^2)$, and using the first Dyson equation to remove $g$, we get
\begin{equation}
	\mathcal{G} + t^2 G_0^2 \mathcal{G} = G_0 + t^2 G_0 G \mathcal{G}.
\end{equation}
Using $G = \mathcal{G} (F + 1)$, we get
\begin{equation}
	\mathcal{G} = G_0 + t^2 G_0 (\mathcal{G} F + \mathcal{G} - G_0) \mathcal{G}.
\end{equation}

Now we derive the order-by-order self-consistency condition.
We want to isolate $\mathcal{G}_N$ and write it as a function of $F_1, \ldots, F_N$ and $\mathcal{G}_0, \ldots, \mathcal{G}_{N-1}$.
We will use $[\ldots]_N$ to denote the $N$th order coefficient of the series enclosed in brackets.

Extracting the order $N > 0$ coefficient gives
\begin{equation}
	\mathcal{G}_N = t^2 G_0 \qty[\mathcal{G} F \mathcal{G}]_N + t^2 G_0 \qty[\qty(\mathcal{G} - G_0)\mathcal{G}]_N
\end{equation}
Using the fact that $F_0 = 0$, the first term on the right-hand side is independent of $\mathcal{G}_N$,
\begin{gather}
	\qty[\mathcal{G} F \mathcal{G}]_N = \qty[\mathcal{G}^{(N-1)} F^{(N)} \mathcal{G}^{(N-1)}]_N,
\end{gather}
while the second is not and reads, after isolating $\mathcal{G}_N$, and using $G_0 = \mathcal{G}_0$:
\begin{equation}
	\qty[\qty(\mathcal{G} - G_0)\mathcal{G}]_N = \mathcal{G}_N \mathcal{G}_0 + \qty[\mathcal{G}^{(N-1)}\mathcal{G}^{(N-1)}]_N.
\end{equation}
Together, we get
\begin{equation}
	\label{eq:sc_bethe_obo_real_times}
	\mathcal{G}_N - t^2 G_0 \mathcal{G}_N G_0 = t^2 G_0 \qty[\mathcal{G}^{(N-1)} \qty(F^{(N)} + 1) \mathcal{G}^{(N-1)}]_N.
\end{equation}

In steady state problems, the retarded component of $g$ at frequency $\omega$ is $g^R(\omega) = 1/(\omega + \mu + i\eta)$, with $\mu$ a chemical potential and $\eta$ a small positive number for regularization.
Retarded components are denoted with a $R$ superscript.
Using Eq.~\eqref{eq:bethe_lattice_dyson}, one can simplify the retarded component of Eq.~\eqref{eq:sc_bethe_obo_real_times} into an explicit expression
\begin{equation}
	\mathcal{G}_N^R = \frac{t^2}{\sqrt{(\omega + \mu)^2 - 4t^2}}\qty[\mathcal{G}^{R(N-1)} \qty(1 + F^{R(N)}) \mathcal{G}^{R(N-1)}]_N
\end{equation}
Note that, from the denominator $\sqrt{(\omega + \mu)^2 - 4t^2}$, we expect singular behavior at the band edges $\omega = -\mu \pm 2t$.

\section{Implementation details}
\label{app:implementation}

In this appendix, we present some implementation details for the results of Sec.~\ref{sec:illustration}.

\paragraph{Impurity solver}
We used the ``full kernel" \ac{QQMC} algorithm of Ref.~\onlinecite{Bertrand2021} to obtain perturbation series of (retarded) $F$ in real times.
We use a maximum integration time $t_{\rm M} = 10^3$.
Very little modifications are needed to the original implementation, as only the integrand is changed. The warping technique and the model function are defined in the same way, with the updated integrand.
We found that the model function is as good at capturing the new integrand as the original one, and we witness an improved convergence scaling when compared to Monte Carlo.

Integration errors are estimated using the standard randomized quasi-Monte Carlo approach~\cite{Dick2013, LEcuyer2018, Macek2020}. In this work we use $N_{\rm sh} = 100$ randomly shifted Sobol' sequences.
We estimate the integration error by $\sigma / \sqrt{N_{\rm sh}}$ with $\sigma$ the standard deviation over the $N_{\rm sh}$ samples.
Propagation of error from one order to the next is not taken into account, but it has been checked that it is negligible compared to the integration error at current order.

\paragraph{Self-consistency}
We apply the self-consistency condition between the retarded $F$ and Weiss function in real frequencies from Eq.~\eqref{eq:sc_bethe_obo}, and derived in App.~\ref{app:sc_bethe_lattice}.
Lesser and greater Weiss functions are obtained from the retarded one using fluctuation-dissipation relations.
The fast Fourier transform algorithm is used to switch back and forth between real times and real frequencies.

\paragraph{Resummation}
The quasiparticle peak heights shown in Fig.~\ref{fig:quasiparticle_height} are obtained from the perturbation series of the spectral weight up to order 6, using Pad\'e approximants.
For all temperatures except the highest ($T/D = 0.147$), the $(2, 1)$ and $(1, 2)$ approximants in the variable $U^2$, are shown and match almost perfectly within the interaction range displayed.
For the highest temperature, we had to fall back to the $(1, 1)$ approximant.

\section{Comparison with analytic results for first orders}
\label{app:comp_exact}

\begin{figure}[t]
	\centering
	\includegraphics[width=\linewidth]{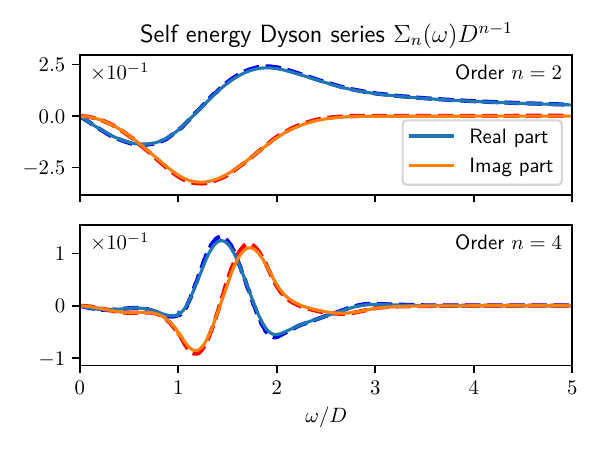}
	\caption{
		\label{fig:self_energy_high_beta}
		Same as in Fig.~\ref{fig:self_energy} but at low temperature $T = D/60$ (plain lines), and exact zero temperature perturbation orders from Ref.~\cite{GebhardNoack_2003} (dashed lines, blue for real part, red for imaginary part).
	}
\end{figure}

Figure~\ref{fig:self_energy_high_beta} shows that the perturbation orders computed with our method match analytical calculations from Ref.~\cite{GebhardNoack_2003}.
In Ref.~\cite{GebhardNoack_2003}, the 2 first non-zero perturbation orders for the self-energy in the half-filled Hubbard model on the Bethe lattice have been calculated analytically at zero temperature.
In Fig.~\ref{fig:self_energy_high_beta} we compare them to the perturbation orders we obtain at low temperature ($T = D/60$), and observe a very good agreement.
We expect that the small difference be explained by both the small mismatch in temperatures and the smoothing parameter $\eta$ that is absent from the calculations of Ref.~\cite{GebhardNoack_2003}.
Note that such analytical calculations are only accessible for very low perturbation orders.

\bibliography{biblio.bib}

\end{document}